\documentclass[aps]{revtex4}
\usepackage{graphicx,psfrag}
\def\bc{\begin{center}}
\def\ec{\end{center}}

\def\beq{\begin{equation}}
\def\eeq{\end{equation}}

\begin{document}

\title{
Quantum transport with strong scattering: \\
Beyond the nonlinear sigma model 
}

\author{K. Ziegler}
\affiliation{Institut f\"ur Physik, Universit\"at Augsburg\\
D-86135 Augsburg, Germany\\
}
\date{\today}

\begin{abstract}
Transport properties of a two-band system with spectral nodes are studied in the presence
of random scattering. Starting from a Grassmann functional integral, we derive a bosonic
representation that is based on random phase fluctuations. Averaging leads to a graphical
representation of the correlation function with entangled random walks and 4-vertices. In 
the strong scattering limit we derive a complex transition amplitude. For the example of
two-dimensional Dirac fermions we obtain a localization length proportional to the inverse 
scattering rate.
\end{abstract}

\maketitle

\section{Introduction}

Diffusion, a concept known from classical physics, where the mean-square displacement of a particle position grows
linearly with time, can also be observed in quantum systems. It provides our basic understanding for a large number 
of transport phenomena, such as the metallic behavior in electronic systems. The diffusion coefficient $D(E)$
of quantum particles of energy $E$ is defined by the correlation of two one-particle Green's functions as 
\cite{thouless74,stone81}
\beq
D(E)=\lim_{\epsilon\to0}\epsilon^2\sum_r  r_k^2\langle G_{r0}(E+i\epsilon)G_{0r}(E-i\epsilon)\rangle_d
\ ,
\label{diff_c0}
\eeq
where $\langle ...\rangle_d$ is the average with respect to a distribution of random scatterers. 
The one-particle Green's function
is defined as the resolvent $G(z)=(H-z)^{-1}$ of the Hamiltonian $H$, and $G_{r0}(E+i\epsilon)$ describes
the propagation of a particle with energy $E$ from the origin to a site $r$.
Scattering is introduced by the fact that the Hamiltonian $H$ is random with mean $\langle H\rangle_d =H_0$.
According to expression (\ref{diff_c0}), a non-vanishing diffusion coefficient requires
a long-range correlation for $\epsilon\sim0$. This can be produced by a spontaneously 
broken symmetry, in analogy to the Goldstone modes in $O(N)$ symmetric field theories. Indeed, $\epsilon$
plays the role of a symmetry-breaking term, such that in the limit $\epsilon\to0$ spontaneous
symmetry breaking is indicated by a non-vanishing scattering rate \cite{wegner79,wegner80}. The corresponding symmetry depends
on the Hamiltonian $H$. A special case appears for a two-band Hamiltonian that is represented in terms
of Pauli matrices as $H=h_0\sigma_0+\cdots +h_3\sigma_3$. Assuming that we can diagonalize all four matrices
$h_0$, ..., $h_3$ by the same unitary transformation $h_j\to\lambda_j$ we get for the spectrum two bands
\beq
E_\pm=\lambda_0\pm\sqrt{\lambda_1^2+\lambda_2^2+\lambda_3^2}
\ .
\label{spectrum00}
\eeq
A spectral node, where the two bands are degenerate, is characterized by $\lambda_1^2+\lambda_2^2+\lambda_3^2=0$, 
such that $\lambda_0$ moves us away from the node at zero energy.

In the presence of random scattering there is also Anderson localization due to strong quantum interference
\cite{anderson58}. In one-band systems this effect is particularly strong in one or two dimensions 
\cite{klein86,froehlich83,leschke05}. However, the situation is different in two-band systems due to interband scattering. 
This causes special effects, such as Klein tunneling, which allows quantum particles to escape from localization.
Transport measurements on graphene support the absence of Anderson localization at least in a weakly disordered
two-dimensional electron gas with spectral nodes \cite{reviews}.

The correlation function of the expression (\ref{diff_c0}) is invariant under non-Abelian chiral transformations if the
Hamiltonian $H$ has a generalized particle-hole symmetry \cite{ziegler12c}
\beq
U H^T U^\dagger =-H , \ \ \ U U^\dagger ={\bf 1}
\label{id1}
\eeq
with a unitary transformation $U$ and the transposition $^T$. 
This can be seen, following Ref. \cite{ziegler12c}, by introducing the Green's function 
${\hat G}(i\epsilon)=({\hat H}+i\epsilon)^{-1}$ for the extended Hamiltonian
\beq
{\hat H}=\pmatrix{
H_+ & 0 & 0 & 0\cr
0 & H_- & 0 & 0\cr
0 & 0 & H_-^T & 0 \cr
0 & 0 & 0 & H_+^T \cr
} , \ \ H_\pm=H\pm\mu\sigma_0
\ ,
\label{8by8a}
\eeq
where $\mu$ is a chemical potential that violates the relation (\ref{id1}) and corresponds to $\lambda_0$
in Eq. (\ref{spectrum00}).
Then, together with property (\ref{id1}), the matrix
\beq
{\hat S}=
\pmatrix{
0 & 0 & \varphi_+ U & 0 \cr
0 & 0 & 0 & \varphi_- U \cr
\varphi_+'U^\dagger & 0 & 0 & 0 \cr
0 & \varphi_-' U^\dagger & 0 & 0\cr
}
\label{symm_tr}
\eeq
for scalar variables $\varphi_j, \varphi_j'$
anticommutes with ${\hat H}$: ${\hat S}{\hat H}=-{\hat H}{\hat S}$. 
This relation implies a non-Abelian chiral symmetry \cite{ziegler12c,ziegler09}:
\beq
e^{\hat S}{\hat H}e^{\hat S}={\hat H}
\label{symmetry1}
\eeq
which is a symmetry relation for the extended Hamiltonian with respect to ${\hat U}=e^{\hat S}$.

Since long-range correlations for the Green's functions in (\ref{diff_c0}) appear due to the
spontaneously broken symmetry (\ref{symmetry1}), it is sufficient to restrict the average $\langle ...\rangle_d$
to an integration with respect to the invariant measure of the symmetry transformation ${\hat U}$.
The invariant measure is related to those degrees of freedom which leave the symmetry 
breaking order parameter invariant (i.e., the scattering rate), as discussed in Ref. \cite{ziegler12c}.
This is similar to an $O(N)$-symmetric Heisenberg ferromagnet with fixed magnetization \cite{coleman}.
Another case is the random-matrix theory, where the invariant measure plays an essential role 
for universal properties such as level spacing \cite{rmt}.
For our model the situation is somewhat different from that of the zero-dimensional random-matrix model, though, 
due to spontaneous symmetry breaking: The invariant measure of the random-matrix theory is defined by
fixing the eigenvalues rather than an order parameter. On the other hand, the order parameter in a $d$-dimensional
system is a spatially fluctuating quantity. Then the determination of the invariant measure requires the approximation 
by a constant order parameter. 

The restriction of the integration with respect to the invariant measure creates a Jacobian, whose leading expansion terms represents
the nonlinear sigma model. The latter provides a conventional approach to study diffusion and the breakdown
of diffusion due to Anderson localization of quantum particles in a disordered environment \cite{wegner79,wegner80}.
For the symmetry (\ref{symmetry1}), however, this gives only diffusion \cite{ziegler09,sinner12}. Therefore, to 
observe Anderson localization in this case we must go beyond the nonlinear sigma model by taking into account
the full Jacobian. 

It should be noticed that the Jacobian factorizes into the $\pm$ sectors as $J=J_+J_-$, 
since the matrices are diagonal with respect to the index according to Eqs. (\ref{8by8a}) and (\ref{symm_tr}).
In the following we consider only one factor and drop the index. 
Then the starting point is the integral with respect to the invariant measure of a non-Abelian chiral symmetry group
in Bose-Fermi space of Ref. \cite{ziegler12c}, where the integration variable is a (complex) Grassmann field $\varphi_r$
with its conjugate field $\varphi_r'$ \cite{berezin,negele}:
\beq
K_{{\bar r}{\bar r}'}
=\frac{1}{\cal N}\int \varphi_{\bar r}\varphi_{{\bar r}'}' J{\cal D}[\varphi],
\ \ \ 
J=\det({\bf 1}+\varphi'\varphi -\varphi'h\varphi h^\dagger)^{-1} , \ \ \ 
{\cal N}=\int J{\cal D}[\varphi]
\ ,
\label{corr00}
\eeq
where
\beq 
h_{rr'}=\sigma_0\delta_{rr'}+ 2i\eta(H_0^T- i{\bar\eta})^{-1}_{rr'}\ \ \ {\rm with}\ {\bar\eta}=\eta+\epsilon
\label{def_h}
\eeq
with
\beq
hh^\dagger
={\bf 1}-4\epsilon(1-\epsilon){\bar\eta}({H_0^T}^2+{\bar\eta}^2)^{-1}
\ .
\label{unitary0}
\eeq
Thus, $h$ is unitary in the limit $\epsilon\to0$. $\eta\ge 0$ is the scattering rate, which
is also the order parameter for spontaneous breaking of the symmetry (\ref{symmetry1}).
In the absence of spontaneous symmetry breaking (i.e. for $\eta=0$), the Jacobian $J$ becomes 1.

The correlation function $K_{{\bar r}{\bar r}'}$ agrees for large distances 
$|{\bar r}-{\bar r}'|$ with the correlation of the Green's functions in (\ref{diff_c0}) 
\[
K_{{\bar r}{\bar r}'}\sim\langle G_{{\bar r}{\bar r}'}(E+i\epsilon)G_{{\bar r}'{\bar r}}(E-i\epsilon)\rangle_d
\]
and can be used to calculate the diffusion coefficient of Eq. (\ref{diff_c0}).
The nonlinear sigma model is obtained from the expansion of the logarithm of the Jacobian $\log J$ up to second order
in the Green's function $2i\eta(H_0^T\pm i{\bar\eta})^{-1}$.
In the following, though, we will not use this approximation but treat the full Jacobian $J$ of Eq. (\ref{corr00}).

\section{Summary of the results}

Starting from a previously formulated fermionic functional integral (\ref{corr00})
of scattered quantum particles in two-band systems 
with nodes \cite{ziegler12c}, we derive a transformation from the fermionic 
(Grassmann) to a bosonic (complex) integral (Sect. \ref{sect:fint}). This is used
to express the scattering of the quantum states in terms of entangled random walks
in Sects. \ref{sect:bint}, \ref{sect:lce}.
The transformation provides eventually the expression
\[
K_{{\bar r}{\bar r}'}
=\langle C^{-1}_{{\bar r}{\bar r}'}\rangle
\equiv\frac{\langle C^{-1}_{{\bar r}{\bar r}'}\det C\rangle_a}{\langle\det C\rangle_a} \ \ \ {\rm with}\ 
C_{rr'}= 2\delta_{rr'}-\sum_{j,j'}e^{i\alpha_{rj}}h_{rj,r'j'}\sum_{j'',r''}h^\dagger_{r'j',r''j''}e^{-i\alpha_{r''j''}}
\ ,
\]
where the brackets $\langle ...\rangle_a$ mean averaging with respect to the angular variables 
$\{\alpha_{rj}\}$ of $C$, as defined in Eq. (\ref{angular_i}). The average $\langle ...\rangle$ is the 
angular average with the additional complex weight $\det C/\langle\det C\rangle_a$. The result of the 
averaging process has a simple graphical interpretation by two types of random walks 
which are entangled by 4-vertices, as illustrated in Figs. \ref{fig:ssa1}, \ref{fig:ssa2}.

The correlation function is treated in Sect. \ref{sect:3-vertex}, which is based on the expansion in powers of $C-\langle C\rangle_a$
around the unperturbed correlation function
\[
g=
({\bf 1}- P_\epsilon)^{-1}
\]
with the transition probability
\[
P_{\epsilon,rr'}  =\delta_{rr'}-\langle C_{rr'}\rangle_a/2
=\sum_{j,j'}|h_{rj,r'j'}|^2  \ \ \ {\rm with}\ \
\lim_{\epsilon\to0}\sum_{r'}P_{\epsilon,rr'}=1
\]
for a fixed time interval. $P_{\epsilon,rr'}$ describes a classical random walk, and from its
Fourier transform ${\tilde P}_{\epsilon,k}$ we get a diffusion pole 
for $g$ at $\epsilon=0$ and $k=0$. Thus, the leading term of this expansion is a diffusion process.
Including the next order term of the expansion (cf. Fig. \ref{fig:ssa3}) there is a complex transition amplitude 
which reads $P=P_\epsilon +i\delta P$  with 
\[
\delta P_{r r'}
=\frac{1}{2}Im\sum_{j,j'}(h{\bar h}^\dagger h)_{rj,r'j'}h^\dagger_{r'j',rj}
\ .
\]
The imaginary correction indicates localization because the pole of the propagator $K=({\bf 1}-P)^{-1}$ has moved away 
from the real axis. A special case exists for two-dimensional Dirac fermions, which is studied in Sect. \ref{sect:strong} 
for strong scattering. This requires a shift of the expansion point.
In contrast to the diffusive case above, the correlation function (\ref{new_prop})
has two poles which are separated by a distance proportional to the scattering rate.

\section{Transformation from fermionic to bosonic integrals}

The starting point for our calculation is the Jacobian in Eq. (\ref{corr00}) \cite{ziegler09} 
\beq
J=\det({\bf 1}+\varphi'\varphi -\varphi'h\varphi h^\dagger)^{-1}
\ ,
\label{jacobian00}
\eeq
where $h_{\pm,r,r'}$ is defined in Eq. (\ref{def_h}).
The calculation will be more transparent after a transformation from the Grassmann integral
to a complex integral. This is based on the fact that the inverse determinant can also be written as an 
integral 
\beq
\det({\bf 1}+\varphi'\varphi -\varphi'h\varphi h^\dagger)^{-1}
=\int\exp\left[-\chi\cdot({\bf 1}+\varphi'\varphi -\varphi'h\varphi h^\dagger){\bar\chi}\right]\prod_{r,j} d\chi_{rj}
\label{fb0}
\eeq
with respect to a two-component complex (bosonic) field $\chi_{rj}$ with spinor index $j=1,2$.

\subsection{Integrating out fermions}
\label{sect:fint}

After replacing the inverse determinant in Eq. (\ref{corr00}) by the right-hand side of (\ref{fb0}) and interchanging
the bosonic integration with the fermionic integration, we can perform the latter
because the fermion field appears only as a quadratic form in the exponent. First, however, we use the identity
\[
\int\varphi_{\bar r}\varphi_{{\bar r}'}'
\exp\left[\chi\cdot(-\varphi'\varphi +\varphi'h\varphi h^\dagger){\bar\chi}\right]\prod d\varphi_{rj}
\]
\[
=\frac{\partial}{\partial\alpha_{{\bar r}'{\bar r}}}
\int\exp(-\varphi_{{\bar r}'}'\alpha_{{\bar r}'{\bar r}}\varphi_{\bar r})
\exp\left[\chi\cdot(-\varphi'\varphi +\varphi'h\varphi h^\dagger){\bar\chi}\right]\prod d\varphi_{rj}
\Big|_{\alpha_{{\bar r}'{\bar r}}=0}
\]
which provides us the following expression
\beq
=\frac{\partial}{\partial\alpha_{{\bar r}'{\bar r}}}\int\exp\left[-\sum_{r,r'}\varphi_{r'}
(C+\alpha)_{r'r}\varphi_{r}\right]\prod_{r,j} d\varphi_{rj}\Big|_{\alpha_{{\bar r}'{\bar r}}=0}
=\frac{\partial}{\partial\alpha_{{\bar r}'{\bar r}}}\det(C+\alpha)\Big|_{\alpha_{{\bar r}'{\bar r}}=0}
=C^{-1}_{{\bar r}{\bar r}'}\det C
\label{det_rep}
\eeq
with the matrix
\beq
C_{rr'}=\sum_j|\chi_{rj}|^2\delta_{rr'}-\sum_{j,j'}\chi_{rj}h_{rj,r'j'}
\sum_{j'',r''}h^\dagger_{r'j',r''j''}{\bar \chi}_{r''j''}
\ .
\label{corr_matr}
\eeq
Thus, we obtain for the correlation function the bosonic integral
\beq
K_{{\bar r}{\bar r}'}=\frac{\partial}{\partial\alpha_{{\bar r}'{\bar r}}}
\log\left[\int\det(C+\alpha)\prod_{r,j}e^{-|\chi_{rj}|^2}d\chi_{rj}\right]
\Big|_{\alpha_{{\bar r}'{\bar r}}=0}
\ .
\label{boson_int1}
\eeq

\subsection{Integrating out bosons}
\label{sect:bint}

We continue with the bosonic integral in (\ref{boson_int1}) and consider the expansion of the determinant:
\[
\int\det(C+\alpha)\prod_{r,j}e^{-|\chi_{rj}|^2}d\chi_{rj}=\sum_\pi (-1)^\pi
\int\prod_{r}(C+\alpha)_{r\pi(r)}\prod_j e^{-|\chi_{rj}|^2}d\chi_{rj}
\ .
\]
With the expression of $C$ in Eq. (\ref{corr_matr}) there is only a single pair of factors 
$\{\chi_{rj},{\bar\chi}_{rj}\}$ for each site $r$ in each term of the expanded determinant. 
Therefore, the $|\chi_{rj}|$ integration contributes only a factor 
\[
\int e^{-|\chi_{rj}|^2}|\chi_{rj}|^2 d\chi_{rj}=2\pi
\]
such that the integration over $|\chi_{rj}|$ leaves a factor of $2\pi$ and only the phase integration 
remains to be performed in the correlation function in Eq. (\ref{boson_int1}):
\beq
\langle ... \rangle_a=\int_0^{2\pi}...\int_0^{2\pi} ...\prod_{r,j}\frac{d\alpha_{rj}}{2\pi}
\ .
\label{angular_i}
\eeq
Then $C_{rr'}$ in Eq. (\ref{corr_matr}) can be written, after replacing $\chi_{rj}$ by the phase factors, as 
\beq
C_{rr'}= 2\delta_{rr'}-\sum_{j,j'}e^{i\alpha_{rj}}h_{rj,r'j'}{\tilde \chi}_{r'j'},
\ \ \ 
{\tilde \chi}_{r'j'}=\sum_{j'',r''}h^\dagger_{r'j',r''j''}e^{-i\alpha_{r''j''}}
\ 
\label{icorr00}
\eeq
which implies for the correlation function in Eq. (\ref{boson_int1}) the expression
\beq
K_{{\bar r}{\bar r}'}
=\frac{\partial}{\partial\alpha_{{\bar r}'{\bar r}}}\log\langle\det(C+\alpha)\rangle_a
\Big|_{\alpha_{{\bar r}'{\bar r}}=0}
=\frac{\langle C^{-1}_{{\bar r}{\bar r}'}\det C\rangle_a}{\langle\det C\rangle_a}
\equiv \langle C^{-1}_{{\bar r}{\bar r}'}\rangle
\ .
\label{gen_func}
\eeq
This result can be understood as if we have a random walk with additional random phases $e^{\pm i\alpha_{rj}}$, 
presented by $C^{-1}_{{\bar r}{\bar r}'}$, since $hh^\dagger\sim {\bf 1}$ for $\epsilon\sim0$.
The random phases are averaged with respect to the complex weight $\det C/\langle\det C\rangle_a$. 
If we ignore the random phases by setting $\alpha_{rj}=0$, which corresponds to the saddle-point approximation 
($\alpha_{rj}=const.$ is a saddle point due to the $U(1)$ symmetry), the result is 
\beq
C_{rr'}= 2\delta_{rr'}-\gamma\sum_{j,j'} h_{rj,r'j'} , \ \ \ \gamma ={\tilde h}_{k=0,jj}=-1+2\epsilon/{\bar\eta}
\ ,
\label{no_phase}
\eeq
which describes a random walk with transition amplitude $T_{rr'}=(\gamma/2)\sum_{j,j'} h_{rj,r'j'}$.

\subsection{Linked cluster expansion}
\label{sect:lce}

To calculate $\langle\det(C+\alpha)\rangle_a$ in Eq. (\ref{gen_func}) we define
\beq
C'_{rr'}=2\delta_{rr'}-C_{rr'}=\sum_{j,j'}e^{i\alpha_{rj}}h_{rj,r'j'}{\tilde \chi}_{r'j'}
\label{hop1}
\eeq
and use the property of the determinant
\[
\det(C+\alpha)=\exp\left\{Tr[\log(C+\alpha)]\right\}
=2^N\exp\left\{Tr[\log({\bf 1}+(\alpha -C')/2)]\right\}
\ .
\]
The exponent on the right-hand side is expanded as
\beq
\det(C+\alpha)=2^N\exp\left\{-\sum_{l\ge1}\frac{1}{2^l l}Tr[(C'-\alpha)^l]\right\}
\ ,
\label{exp1d}
\eeq
where the trace term in the exponential function
\[
A=-\sum_{l\ge1}\frac{1}{2^l l}Tr[(C'-\alpha)^l]
=-\sum_{l\ge1}\frac{1}{2^l l}\sum_{r_1,r_2,...,r_l}(C'-\alpha)_{r_1r_2}\cdots(C'-\alpha)_{r_lr_1}
\]
can be understood as a summation of clusters: For a given cluster of sites ${\cal C}_l=\{r_1,r_2,...,r_l\}$
we define $a_{{\cal C}_l}=(C'-\alpha)_{r_1r_2}\cdots(C'-\alpha)_{r_lr_1}$, such that the series reads
\[
\sum_{l\ge1}\frac{1}{2^l l}Tr[(C'-\alpha)^l]=\sum_{l\ge1}\frac{1}{2^l l}\sum_{{\cal C}_l}a_{{\cal C}_l}
\ .
\]
In the next step we average the determinant with respect to the phase angles and write
\beq
\langle e^A\rangle_a=\exp\left(\sum_{m\ge1}p_m\right)
\ ,
\label{lct}
\eeq
where $\{ p_m\}$ are linked (or truncated) correlations \cite{glimm}, which can be expressed as
\beq
p_m=\frac{1}{m!}\frac{\partial^m}{\partial\lambda ^m}\left(\log\langle e^{\lambda A}\rangle_a\right)
\Big|_{\lambda=0}
=\frac{1}{m!}\frac{\partial^{m-1}}{\partial\lambda ^{m-1}}\frac{\langle A e^{\lambda A}\rangle_a}{\langle 
e^{\lambda A}\rangle_a}\Big|_{\lambda=0}
\ .
\eeq
The latter is implied by the Taylor expansion of $f(A)$ around $A=0$:
\[
f(A)=\sum_{m\ge 0}\frac{A^m}{m!}\frac{\partial^m}{\partial A^m}f(a)\Big|_{A=0}
=\sum_{m\ge 0}\frac{A^m}{m!}\frac{\partial^m}{\partial (\lambda A)^m}f(\lambda A)\Big|_{\lambda=0}
=\sum_{m\ge 0}\frac{1}{m!}\frac{\partial^m}{\partial \lambda^m}f(\lambda A)\Big|_{\lambda=0}
\ .
\]
Examples are $p_0=0$, $p_1=\langle A\rangle_a$, $p_2=(\langle A^2\rangle_a-\langle A\rangle_a^2)/2$ etc.
In simple words, factorizing products 
$\langle a_{{\cal C}_1}a_{{\cal C}_2}...\rangle_a=\langle a_{{\cal C}_1}...\rangle_a\langle a_{{\cal C}_2}
...\rangle_a$
do not contribute to $p_m$.

From Eq. (\ref{gen_func}), together with the exponential representation (\ref{exp1d}) and the linked cluster 
relation (\ref{lct}), we obtain for the correlation function the linked cluster expansion
\beq
K_{{\bar r}{\bar r}'}=\langle C^{-1}_{{\bar r}{\bar r}'}\rangle
\equiv\frac{\langle C^{-1}_{{\bar r}{\bar r}'}\det C\rangle_a}{\langle\det C\rangle_a}
=\frac{\partial}{\partial\alpha_{{\bar r}'{\bar r}}}\log\langle e^A\rangle_a\Big|_{\alpha_{{\bar r}'{\bar r}}=0}
=\sum_{m\ge1}\frac{\partial p_m}{\partial\alpha_{{\bar r}'{\bar r}}}\Big|_{\alpha_{{\bar r}'{\bar r}}=0}
\ ,
\label{corr4}
\eeq
where the right-hand side can be represented graphically by two types of entangled random walks forming loops.
This is described in the next section.

\subsection{Interpretation and graphical representation}

\begin{figure}
\begin{center}
\includegraphics[width=6cm,height=5cm]{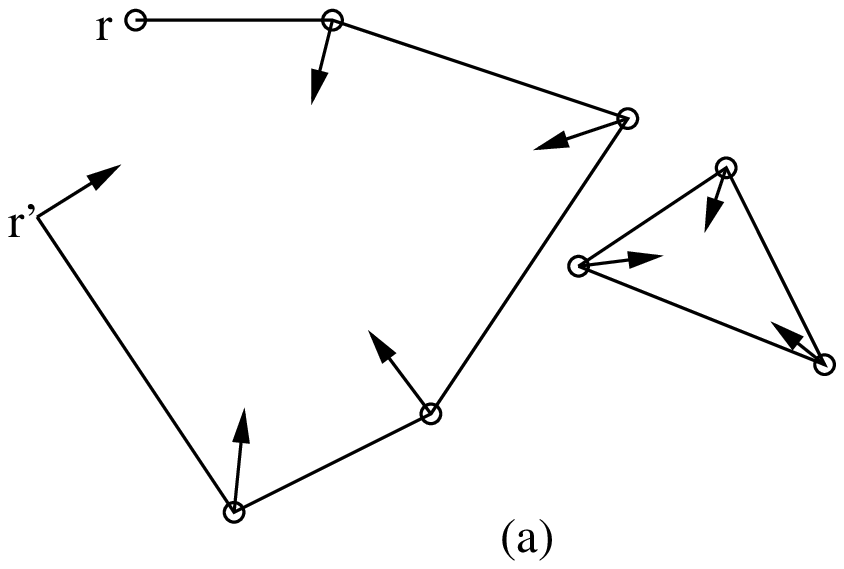}
\includegraphics[width=6cm,height=5cm]{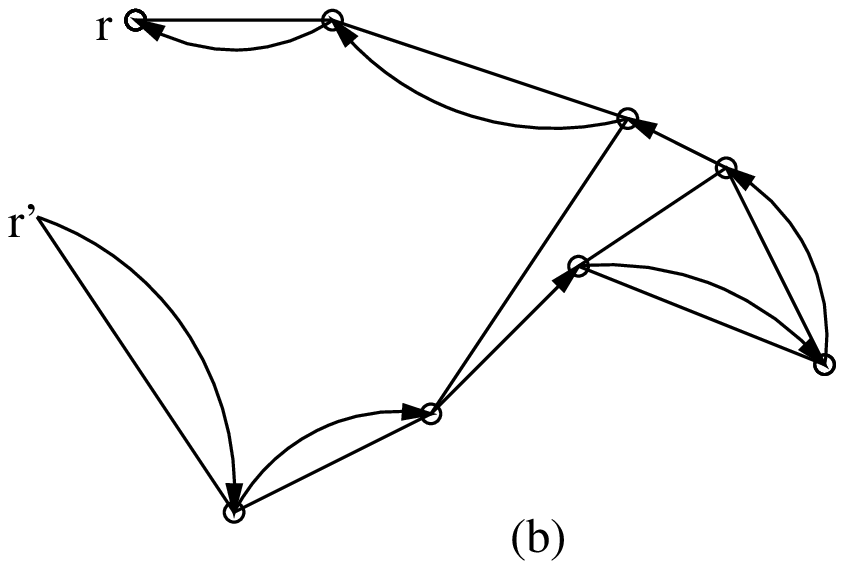}
\caption{
Examples for the graphical representation of the correlation function $K_{rr'}$ with an extra loop 
from the contribution of the determinant in Eq. (\ref{corr4}): (a) before angular averaging, 
(b) a linked cluster after angular averaging.
}
\label{fig:ssa2}
\end{center}
\end{figure}

A cluster contribution to the expression (\ref{corr4}) describes either a loop connecting the cluster 
sites $r_1,...,r_l$
\beq
a_{\cal C}|_{\alpha_{{\bar r}'{\bar r}}=0}
=\frac{1}{2^l l}C'_{r_1r_2}\cdots C'_{r_lr_1}
\label{prod1}
\eeq
or a corresponding open walk which starts at ${\bar r}'$ and terminates at ${\bar r}$ 
if $r_k={\bar r}'$, $r_{k+1}={\bar r}$:
\beq
\frac{\partial a_{\cal C}}{\partial\alpha_{{\bar r}'{\bar r}}}\Big|_{\alpha_{{\bar r}'{\bar r}}=0}
=\frac{1}{2^l l}C'_{r_1r_2}\cdots C'_{r_{k-1}{\bar r}'}C'_{{\bar r},r_{k+2}}\cdots C'_{r_l r_1}
\ ,
\label{prod2}
\eeq
where each step between sites $r$ and $r'$ contributes the (complex) weight $C'_{rr'}$ of (\ref{hop1}).
Then a linked cluster of Eq. (\ref{corr4}) is a product of an open walk and $n$ ($n=0,1,...$) loops, 
which are connected by the angular integration. 
Both, the loop and the open walk, are graphically represented in Figs. \ref{fig:ssa1} (a) and \ref{fig:ssa2} (a). 
The circles represent the phase factors $e^{i\alpha_{rj}}$ and the arrows represent 
${\tilde \chi}_{r'j'}=\sum_{j'',r''}h^\dagger_{r'j',r''j''}e^{-i\alpha_{r''j''}}$.
Now we follow the sites with phase factors. The cancellation of the latter results in a random
walk consisting of alternating steps $h$ and $h^\dagger$, which is visualized by connecting the 
arrows with the circles.
Then we connect the intermediate sites without a phase factor with sites with a phase factor 
$e^{i\alpha_{rj}}$ to take care of the site $r_j$ in $C_{r_{j-1}r_j}'C_{r_j r_{j+1}}'$. 
This procedure connects all circles and arrows pairwise and creates 4-vertices that are connected 
either by $h$ or by $h^\dagger$ steps (cf. Figs. \ref{fig:ssa1} (b), (c) and \ref{fig:ssa2} (b)). 
The linked cluster property is reflected by the fact that the loops and the open walk are 
connected (linked) by $h^\dagger$ steps (cf. Fig. \ref{fig:ssa2} (b)). 
The formation of 4-vertices can be seen as a self-crossing which is enforced by the angular integration. 

There is an alternative interpretation of the graphs by two types of walks with either $h$ or with $h^\dagger$ 
steps only. This can be seen as that the $h$ walk from ${\bar r}'$ to ${\bar r}$ is 
decorated with the walk consisting of $h^\dagger$ steps. Thus, the averaging of phase factors leads to 
a decoration of the walk with $h$ steps. This decoration renormalizes the
diffusion coefficient (i.e., the coefficient of the off-diagonal elements of $h$). Here it should be noticed 
that any additional crossing of these walks does not contribute an extra weight,
implying that the decorated walks are not interacting.

The number of graphs to each order of the 4-vertex grows exponentially due to the random-walk nature
of the expansion, connecting the two sites of the correlation function. Therefore, the expansion 
converges for a sufficiently small contribution to each connection between two 4-vertices. 


\subsection{3-vertex expansion}
\label{sect:3-vertex}

\begin{figure}
\begin{center}
\includegraphics[width=5cm,height=5cm]{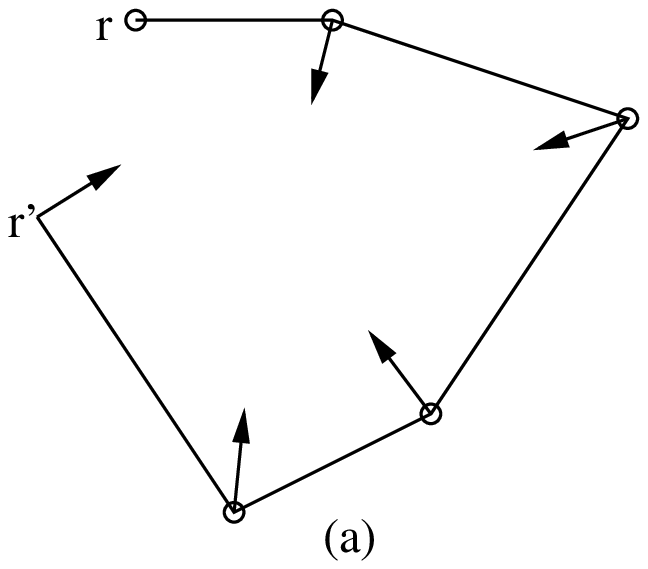}
\includegraphics[width=5cm,height=5cm]{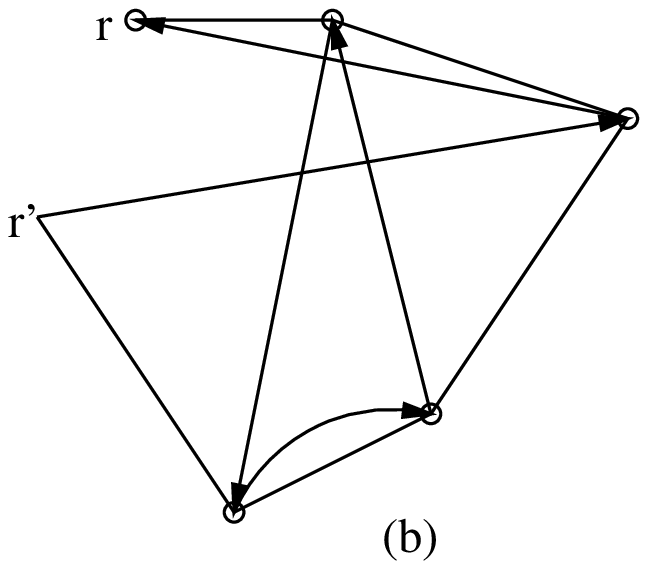}
\includegraphics[width=5cm,height=5cm]{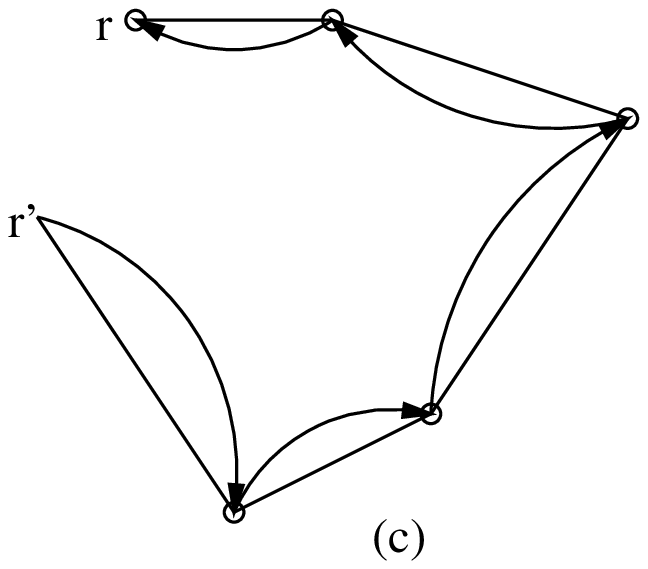}
\caption{Examples for the graphical representation of the correlation function $K_{rr'}$ from Eq. (\ref{corr4}):
(a) before angular averaging, (b), (c) two special cases after angular averaging. The latter leads
to a pairwise connection of the arrows and the circles in (a), which creates 4-vertices.
}
\label{fig:ssa1}
\end{center}
\end{figure}

A shift of the expansion point in Eq. (\ref{exp1d}) by $\langle C'\rangle_a$ means that we take into
account the summation over terms from $\langle C'\rangle_a$ in leading order of the expansion. An typical
example of such terms is shown in the graph of Fig. \ref{fig:ssa1} (c). The shift can be achieved by using
\beq
\det(C+\alpha)=\frac{2^N}{\det g}\exp\left\{Tr[\log({\bf 1}+\frac{1}{2}g(-C'+\langle C'\rangle_a))]\right\}
\ \ \ {\rm with} \ \ g=\left({\bf 1}+\frac{1}{2}\alpha -\frac{1}{2}\langle C'\rangle_a\right)^{-1}
\eeq
and employing the expansion in powers of $\delta C=C'-\langle C'\rangle_a$
\beq
Tr[\log({\bf 1}-\frac{1}{2}g\delta C)]=-\sum_{l\ge1}\frac{1}{2^l l}Tr\left[(g\delta C)^l\right]
\ .
\label{expansion3}
\eeq
Then the leading order of the correlation function reads
\beq
K\approx
({\bf 1}- P_\epsilon)^{-1} 
\label{diff_prop}
\eeq
with the transition probability for a fixed time interval
\beq
P_{\epsilon,rr'}=\langle C'_{rr'}\rangle_a/2=\sum_{j,j'}|h_{rj,r'j'}|^2 , \ \ \ 
\lim_{\epsilon\to0}\sum_{r'}P_{\epsilon,rr'}=1
\ .
\label{class_rw}
\eeq
This approximation agrees with the nonlinear sigma model of Ref. \cite{ziegler12c} and describes diffusion.
The graphical representation of the expansion to higher orders is characterized by a walk with alternating
steps of $h$ and $h^\dagger$ and by 3-vertices, which are connecting pairwise points of 
the alternating walk with the diffusion propagators $g$. An example is depicted in Fig. \ref{fig:ssa3}. 
Including this correction yields for the correlation function
\beq
K_{{\bar r}'{\bar r}}\sim g_{{\bar r}'{\bar r}}+\frac{1}{4}\left[g\left(
\langle\delta C g\delta C\rangle_a-\langle \delta C Tr(g\delta C)\rangle_a\right)g\right]_{{\bar r}'{\bar r}}
\label{p_correct1}
\eeq
\[
\sim \left( g^{-1}-\frac{1}{4}\left(
\langle\delta C g\delta C\rangle_a-\langle \delta C Tr(g\delta C)\rangle_a\right)\right)^{-1}_{{\bar r}'{\bar r}}
\]
if the correction is small. The two terms inside the brackets are those in Fig. \ref{fig:ssa3}.
This correction can be understood in terms of the expansion that the transition probability $P_\epsilon$ in Eq. (\ref{diff_prop})
is replaced by  $P=P_\epsilon+i\delta P$ with the correction term (details in App. \ref{app:pt})
\beq
\delta P_{r r'}
=\frac{1}{2}Im\sum_{j,j'}(h{\bar h}^\dagger h)_{rj,r'j'}h^\dagger_{r'j',rj} \ \ \ 
{\rm with}\ \ {\bar h}_{rj,r'j'}=g_{rr'}h_{rj,r'j'} 
\ .
\label{p_correct2}
\eeq
Thus, we have replaced the classical random walk in Eq. (\ref{class_rw}) by a quantum random walk, 
where the transition probability $P_\epsilon$ has been replaced by a complex transition amplitude 
$P=P_\epsilon+i\delta P$. 
Since  $\sum_{r'}\delta P_{rr'}=0$, the completeness of the transition 
probability $\sum_{r'}P_{\epsilon,rr'}=1$ is preserved. However, the pole of the correlation function $K$
with respect to $\epsilon$ is not at $\epsilon=0$ any more but has moved away from the real axis.
It depends on the details of $h$ whether or not localization appears. An example is discussed in
the next section. 

\section{Strong scattering limit for two-dimensional Dirac fermions}
\label{sect:strong}

Starting from the expression (\ref{def_h}), we can estimate the terms of the 3-vertex expansion.
First, it should be noticed that $\delta C$ is of order $1/{\bar\eta}$. The properties of the expansion
depend on the specific choice of the Hamiltonian $H_0$. In the case of two-dimensional 
Dirac fermions we have $H_0={\vec\sigma}\cdot{\vec k}=k_1\sigma_1+k_2\sigma_2$ with the momentum cut-off $\lambda$:
$k\le\lambda$. 
This gives $g_k\sim {\bar \eta}^2/(\epsilon/{\bar \eta}+k^2/2\pi)$, which diverges like ${\bar\eta}^2$. Thus, in this case
the 3-vertex expansion cannot be controlled by powers of $1/{\bar\eta}$. However, this problem is fixed by shifting
the expansion point $\langle C'\rangle_a \to {\bar C}$, where we replace the random phase factors in $C'$ of (\ref{hop1}) 
by a constant angle $\alpha_{rj}=\phi_j$: 
\beq
{\bar C}_{rr'}=-\sum_{j,j'}h_{rj,r'j'}e^{i(\phi_j-\phi_{j'})}
\eeq
whose Fourier components read
\beq
{\bar C}_k = - 2\left(1-\frac{2\eta{\bar\eta}}{k^2+{\bar\eta}^2}+\frac{4i\eta}{k^2+{\bar\eta}^2}{\vec k}\cdot{\vec s}
\right)
\eeq
with the two-component unit vector ${\vec s}=(\cos(\phi_1-\phi_2),\sin(\phi_1-\phi_2)$. For $\eta\gg\lambda$ we get
\beq
{\bar C}_k\sim 2\left(1-\frac{2\epsilon}{{\bar\eta}}-\frac{4i}{\bar\eta}{\vec k}\cdot{\vec s}
-\frac{2k^2}{{\bar\eta}^2}\right)
\ .
\eeq
The determinant $\det({\bf 1}-{\bar C}/2)$ does not depend on the global phase difference $\phi_1-\phi_2$,
which reflects the invariance of the angular integration under a global rotation. 

A consequence of the shift is that we now have
\beq
Tr\{\log[{\bf 1}-\frac{1}{2}\gamma(\delta C +\langle C'\rangle_a -{\bar C})]\}=-\sum_{l\ge1}\frac{1}{2^l l}
Tr\left\{[\gamma(\delta C +\langle C'\rangle_a -{\bar C})]^l\right\}
\label{expansion4}
\eeq
instead of the expansion (\ref{expansion3}) with
\beq
\gamma=\left({\bf 1}+\frac{1}{2}\alpha -\frac{1}{2}{\bar C}\right)^{-1}
\ .
\eeq
The latter is of order ${\bar\eta}$, since ${\bar C}$ is of order ${\bar\eta}^{-1}$. 

\begin{figure}
\begin{center}
\includegraphics[width=10cm,height=5cm]{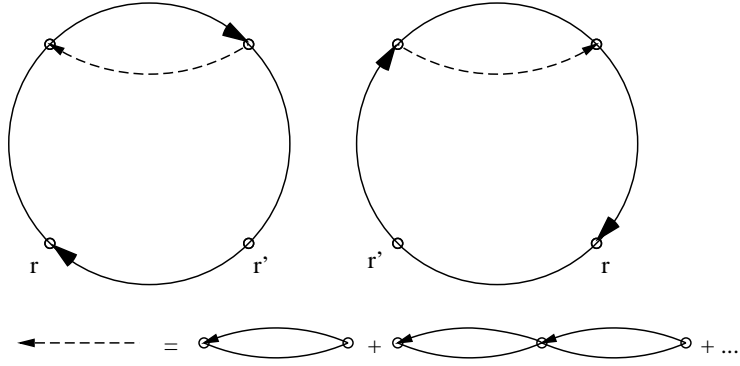}
\caption{
The two terms of the leading correction (\ref{p_correct1}) in the 3-vertex expansion. The dashed line
is the diffusion propagator $g$ of Eq. (\ref{diff_prop}).
}
\label{fig:ssa3}
\end{center}
\end{figure}

In comparison with the expansion (\ref{expansion3}) we have extra terms $\gamma(\langle C'\rangle_a -{\bar C})$
in the expansion (\ref{expansion4}). These terms can be collected as a geometric series and lead to factors $g$;
i.e., they replace $\gamma\to g$.
On the other hand, they must be combined with other terms of the expansion which are of the same order.
For instance, the correction of Fig. \ref{fig:ssa3} has additional terms, which are generated by replacing
$\delta C$ with $\langle C'\rangle_a -{\bar C}$.

The leading term of the expansion
\beq
K_q\approx\frac{1}{1-\frac{1}{2}{\bar C}_q}
\sim\frac{{\bar\eta}/2}{\epsilon +2i{\vec q}\cdot{\vec s}+q^2/{\bar\eta}}
\label{new_prop}
\eeq
is not a diffusion propagator but it has the two separated poles 
$q=-i{\bar\eta}\cos\varphi\pm i\sqrt{{\bar\eta}\epsilon+{\bar\eta}^2\cos^2\varphi}$, where $\varphi$ is the angle
between ${\vec q}$ and ${\vec s}$. For a given ${\vec q}$ we can always choose ${\vec s}$ such that $\varphi=0$.
The diffusion propagator, on the other hand, has the poles $q=\pm i\sqrt{{\bar\eta}\epsilon}$. This difference in the
pole structure is crucial when we apply a Fourier transformation to real space, since the diffusion propagator forces 
the path of the $q$--integration to 0 for $\epsilon\to0$, whereas the propagator (\ref{new_prop}) allows the path of the
$q$--integration to move away from the real axis by the distance $2{\bar\eta}$.
This implies an intrinsic length scale $\xi\propto{\bar\eta}^{-1}$ which represents a localization length.
Moreover, $K_0={\bar\eta}/2\epsilon$ diverges with $\epsilon\sim 0$, as it is required by the identity
\beq
K_0\approx \sum_r \langle G_{r0}(E+i\epsilon)G_{0r}(E-i\epsilon)\rangle_d
=\frac{i}{2\epsilon}\langle G_{00}(E+i\epsilon)-G_{00}(E-i\epsilon)\rangle_d
\ .
\eeq

\appendix

\section{Perturbation Theory}
\label{app:pt}

Together with
\[
\langle \delta C_{r_1r_2}\delta C_{r_3r_4}\rangle_a
=\sum_{j_1,...,j_4}h_{r_1j_1,r_2j_2}h^\dagger_{r_2j_2,r_3j_3}h_{r_3j_3,r_4j_4}h^\dagger_{r_4j_4,r_1j_1}
\]
we obtain from (\ref{p_correct1}) the expression
\[
g_{{\bar r}'{\bar r}}
+\frac{1}{4}\sum_{r_1}g_{{\bar r}'r_1}\sum_{r_2,r_3,r_4}g_{r_2r_3}\sum_{j_1,...,j_4}
h_{r_1j_1,r_2j_2}h^\dagger_{r_2j_2,r_3j_3}h_{r_3j_3,r_4j_4}h^\dagger_{r_4j_4,r_1j_1}
g_{r_4{\bar r}}
\]
\[
-\frac{1}{4}\sum_{r_1}g_{{\bar r}'r_1}\sum_{r_3,r_2,r_4}g_{r_4r_3}\sum_{j_1,...,j_4}
h_{r_1j_1,r_2j_2}h^\dagger_{r_2j_2,r_3j_3}h_{r_3j_3,r_4j_4}h^\dagger_{r_4j_4,r_1j_1}g_{r_2{\bar r}}
\ .
\]
The corresponding graphs are depicted in Fig. \ref{fig:ssa3}. This leads to a correction
of the transition probability of the classical walk in Eq. (\ref{class_rw}):
\[
\delta P_{r_1 r_2}
=\frac{-i}{4}\sum_{r_2,r_3}g_{r_2r_3}\sum_{j_1,...,j_4}
h_{r_1j_1,r_2j_2}h^\dagger_{r_2j_2,r_3j_3}h_{r_3j_3,r_4j_4}h^\dagger_{r_4j_4,r_1j_1}
\]
\[
-\frac{-i}{4}\sum_{r_3,r_2}g_{r_2r_3}\sum_{j_1,...,j_4}
h_{r_1j_1,r_4j_4}h^\dagger_{r_4j_4,r_3j_3}h_{r_3j_3,r_2j_2}h^\dagger_{r_2j_2,r_1j_1}
\ .
\]
This means that 
$\sum_{r'}P_{rr'}=1$ for $\epsilon\to0$. With ${\bar h}_{rj,r'j'}=g_{rr'}h_{rj,r'j'}$ this can also be 
written as 
\[
\delta P_{r_1 r_2}
=\frac{-i}{4}\sum_{j_1,j_2}\left[(h{\bar h}^\dagger h)_{r_1j_1,r_2j_2}h^\dagger_{r_2j_2,r_1j_1}
-\{(h{\bar h}^\dagger h)_{r_1j_1,r_2j_2}h^\dagger_{r_2j_2,r_1j_1}\}^*\right]
\ .
\]
Inside the square brackets the matrix elements of the two terms are complex conjugate to each other,
which implies imaginary matrix elements:
\beq
\delta P_{r_1 r_2}
=\frac{1}{2}Im\sum_{j_1,j_2}(h{\bar h}^\dagger h)_{r_1j_1,r_2j_2}h^\dagger_{r_2j_2,r_1j_1}
\ .
\eeq

\end{document}